# Foundation q-rules

Richard Mould[1]

**Abstract**

The *q-rules* are three auxiliary rules that guide the application of Schrödinger's equation. They are a set of instructions that describe how stochastic choices cause the wave to collapse and "start over" with new boundary conditions.

## Introduction

The Born interpretation of quantum mechanics gives the probability that a system will be found in a certain state at a time $t$ after the initial time $t_0$. In contrast, q-rule equations give a running account of the probability that there will be a stochastic choice during a time interval $dt$, followed by a collapse of the wave function. The Born rule is concerned with the *result* at some time $t$, and the q-rules are concerned with the *process* that leads to a collapse. These two approaches are equivalent in some ways, but the q-rules make a different empirical connection as will be shown. They also have an advantage for reasons having to do with the status of observers in quantum mechanics, and with the possibility of objectifying microscopic physics.

The q-rules allow the *primary* quantum mechanical observer to be included in the system. This is similar to classical physics in that an observer who investigates an external system has the option of extending the system to include himself, thereby allowing him to objectively describe his own experience from moment to moment. Standard quantum mechanics does not let that happen. The Born rule supposes that the primary observer remains outside the system. He may peek at the system from time to time to determine the Born connection at any moment, but he cannot *follow* internal processes. On the other hand, the q-rules equations follow internal processes. This difference has to do with the different ways that probability is introduced – as a result or a process. Probability according to the Born rule refers to an ensemble of trials that derive from a theory that makes no ontological claims. Internal mathematical processes are said *not* to follow internal physical process. On the other hand the q-rules support an internal ontology that embraces probability in individual cases as part of an objective process. The theory claims to describe the physics at all levels.

---

[1] Department of Physics and Astronomy, State University of New York, Stony Brook, New York 11795-3800. http://ms.cc.sunysb.edu/~rmould

Another consequence of the q-rules is that all *secondary* observers can be included in a system in an unambiguous way. The q-rules remove the paradox associated with Schrödinger's cat experiment and with all other ambiguities that result when a secondary conscious observer is admitted into the system. Under the q-rules *all* observers have an unambiguous place in quantum mechanics, as they do in classical physics.

## Some definitions

If a single component of a quantum mechanical wave function includes all the particles in an isolated system and is not the just one component of an expansion in some representation, and if all the variables except time are integrated out, then we say that it is a *q-rule component* of that system. If the system consists of several distinct parts like an atom *a*, an elementary particle *p*, and a macroscopic instrument *m*, then the q-rule component containing these parts is written

$$\psi(t) = apm \otimes E(t)$$

where *E* includes all those parts of the system that are not under consideration and is called the *environment*. If this equation is written as

$$\psi(t) = apmE(t)$$

then the environment may or may not be included as specified.

If the system's wave function $\Psi$ is the sum of two components, then the *q-rule equation* of the total system is written as the sum of the two q-rule components.

$$\Psi(t) = \psi_a(t) + \psi_b(t) \qquad (1)$$

which is normalized to 1.0. The square modulus of the separate components as a function of time is determined by the dynamic principle.

Suppose the component $\psi_b$ in Eq. 1 is initially equal to zero but increases in time as $\psi_a$ decreases due to an interaction. The Schrodinger equation insures that the total square modulus is preserved. In any *quantum jump* it is required that the final component $\psi_b$ either contains a new particle or that it annihilates a particle from $\psi_a$. The other condition implicit in a quantum jump is that the interaction is *non-periodic*. That is, it does not oscillate with a characteristic frequency. This restriction is written into the first q-rule.

Probability is introduced in these rules only as it appears in the second q-rule given below – as related to probability current rather than to square modulus. The *probability current J* into or out of a component refers to a change of its magnitude (i.e., its square modulus) per unit time. We do not give a physical meaning to the square modulus, only to changes in that quantity. The understanding is that probability applies to *individual trials* rather than just to ensembles of trials. Q-rule equations like Eq. 1 refer to individual processes.

We distinguish between *ready* components and *realized* components, where only realized components are understood to have empirical significance. This will be clarified in the examples to be given. Ready components are underlined throughout, whereas realized components appear without an underline.

## The q-rules

The first q-rule describes how ready components are introduced into equations. If a component is not ready then it is realized.

**Q-rule (1)**: *If a non-periodic interaction results in a current flow to a new component that includes a newly created particle or annihilates a previously existing particle, then the new component will be a 'ready' component and will remain so unless converted by q-rule (3).*

[**note:** If the components in Eq. 1 satisfy the required non-periodic and creation or annihilation conditions, then the first one is a 'realized' component that is empirically real, and the second one is 'ready' and is not empirically real. The interaction is therefore written in the form $\Psi(t) = \psi_a(t) + \underline{\psi_b(t)}$ with the second component underlined.]

[**note:** The term "interaction" includes the spontaneous creation of a new component that happens because it is permitted by the dynamic principle.]

______________________________

The second q-rule establishes the existence of a stochastic 'trigger' and identifies a ready component as the 'target' of its stochastic choice.

**Q-rule (2)**: *A systemic stochastic trigger can only strike a ready component, and it does so with a probability per unit time equal to the positive probability current $J$ flowing into it from a realized component.*

[**note:** The target of stochastic choice is a ready component that might be microscopic or macroscopic. That distinction is of no importance. The collapse mechanism does not select a proton or a measuring device as do other theories. Instead it selects non-periodic and creation or annihilation quantum jumps for state reduction.]

______________________________

The collapse of a wave is given by q-rule (3)

**Q-rule (3)**: *When a ready component is stochastically chosen it will begin a transition to a normalized realized component, and all other components will go immediately to zero.*

**[note:** The collapse initiates a transition in which the initial realized state $\psi_a(t)$ in Eq. 1 is replaced by a *realized* state $\psi_b(t)$. This takes place during a time $\Delta T$ that is generally ignored in this paper. In most cases we will speak as though $\psi_a(t)$ goes immediately to $\psi_b(t)$ together with the collapse of other components. A detailed discussion of how the dynamic principle together with and the q-rules produce these transitory results is given in the Appendix.]

**[note:** When the ready component is stochastically chosen it immediately becomes a realized component. It cannot linger between being empirically real and non-real. There is no in-between physical existence and non- existence.]

**[note**: Normalization is automatic because the total state function $\Psi(t)$ in q-rule equations is always equal to 1.0.]

## A particle capture

This first application of the q-rules involves an elementary particle that is captured by a detector.

Apply Schrödinger's equation to a particle $p$ interacting with a detector $d$. The interaction beginning at time $t_0$ is given by the q-rule equation.

$$\Psi(t \geq t_0) = pd_0(t) + \underline{d}_1(t) \qquad (2)$$

where the second component is zero at $t_0$ and increases in time. The free particle $p$ here interacts with the ground state detector $d_0$ producing a probability current flow from the first component to the second, where the latter is the ready detector

$\underline{d}_1$ in its capture state. The interaction is non-periodic. Also, the second component contains particles that do not exist in the first component, including all the cascade photons that are produce in the ion chamber when the particle $p$ enters the chamber. Therefore, the gap given by the + sign satisfies q-rule (1), making $\underline{d}_1(t)$ a 'ready' component as indicated by the underline. Each component in Eq. 2 is assumed to be multiplied by the associated environment (not shown).

Since positive probability current flows into the ready component, it is subject to a stochastic hit as specified by q-rule (2). If that happens at a time $t_{sc}$, then q-rule (3) will require a state reduction giving the q-rule equation

$$\Psi(t \geq t_{sc} > t_0) = d_1(t) \tag{3}$$

The ready component $\underline{d}_1(t)$ in Eq. 2 will also be called the *launch component* because it provides a launch into the realized component $d_1(t)$ in Eq. 3. There will be an initial fuzziness in $d_1(t)$ due to the transient energy $\Delta E$. The interaction that is initiated at $t_{sc}$ extends for the transient time $\Delta T$.

This example shows how a theory based on probability current contrasts with standard quantum mechanics that is based on the Born interpretation. In the Born case, the theory gives the probability that the system will be found in a certain state at a time $t$ after the initial conditions are established at $t_0$. In the q-rule case the theory provides a running account of the probability that a stochastic choice will occur during a time interval $dt$ after $t_0$. This is the reason for the distinction between ready components and realized components. The first component $pd_0$ in Eq. 2 is 'realized' (i.e., its an empirical reality) *until* the stochastic hit at $t_{sc}$, at which time the second component becomes realized in its place. Before that time the second component $\underline{d}_1$ has no empirical significance.

It is possible that the particle will not be captured by the detector, in which case Eq. 2 will not collapse to Eq. 3. The first component $pd_0$ will then continue to be realized and the second component will become irrelevant.

There is an important caveat to this description. The detector has two different kinds of variables: Those that are affected by a stochastic hit and those that are not. It is sometimes important to make this distinction because the Schrödinger equation makes the distinction. A general way of handling this difference is described and its application to detector variables is given in the Appendix that is concerned with *faux processes*.

## Free neutron decay

This section is the first application of q-rules to a microscopic system. The rules are here projected into a realm in which the resulting ontology is discernable, although of course it is non-classical.

*A* free neutron decay is given by the q-rule equation

$$\Psi(t \geq t_0) = n(t) + \underline{e}p\bar{v}(t)$$

where $n$ is a neutron, $e$ is an electron, $p$ is a proton, and $\bar{v}$ is an antineutrino. The second component is zero at $t_0$ and increases in time. It is a ready component, although it not necessary to underline the entire component – one state will do. As before, the launch state $\underline{e}p\bar{v}(t)$ is not empirically real prior to decay. Probability current will flow from the first component to the second, leading to an eventual stochastic hit at time $t_{sc}$. The result is a state reduction given by

$$\Psi(t \geq t_{sc} > t_0) = ep\bar{v}(t)$$

ignoring the transient time $\Delta T$. Assume that the neutron moves across the laboratory in a wave packet of finite width, where the launch component $\underline{e}p\bar{v}(t)$ coincides with the neutron as it goes along. At the time $t_{sc}$ of a stochastic hit, the equation $\Psi(t \geq t_0)$ will collapse and a new solution $\Psi(t \geq t_{sc} > t_0)$ will be launched in its place with initial conditions given by the newly realized component $ep\bar{v}(t_{sc})$. The decay products $e$, $p$, and $\bar{v}$ are now empirically real.

Specific directions of a decay particle's release are not stochastically chosen by this reduction. For that to be determined a detector in the specified direction must be activated. That will require another stochastic hit on the detector given by another q-rule equation.

As in the detector case there are two different kinds of variables associated with the free neutron: Those that are affected by a stochastic hit and those that are not. As before, the general way of handling this difference is developed and its application to this case is given in the Appendix.

## Serial discontinuities

Another macroscopic example considers a counter $C$ that is activated by a nearby radioactive source.

Let $C_0$ mean that no particles have been captured from the radioactive source, let $C_1$ mean that one particle has been captured, and let $C_2$ mean that two particles have been captured, etc., where the series of components $C_0$, $C_1$, $C_2$, $C_3$, … are sequentially connected by non-periodic quantum jumps inasmuch as each includes more cascade photons than are present in the previous component.

A series of captures like this is given by the q-rule equation

$$\Psi(t \geq t_0) = C_0(t) + \underline{C}_1(t) + \underline{C}_2(t) + \underline{C}_3(t) + \ldots \qquad (4)$$

where only $C_0$ is non-zero at time $t_0$. The other components gain amplitude by virtue of probability current flowing first from $C_0$ to $\underline{C}_1$, then to $\underline{C}_2$, and then to $\underline{C}_3$, etc, where the magnitudes of the components form a pulse that moves from left to right in Eq. 4. We do not include the intermediate particle field in this equation because nothing of substance is changed by imagining that the counter components interact directly. Probability current will generally flow into more than one of these components at a time, so current might flow simultaneously into both $\underline{C}_1$ and $\underline{C}_2$, suggesting that $\underline{C}_2$ might be stochastically chosen before $\underline{C}_1$. That is a very unphysical result because a counter cannot record the capture of two particles before it has recorded the captured one particle. The second q-rule insures that that does not happen. It says that the stochastic trigger will only strike when current flows into a ready component *from a realized component.* This means that only $\underline{C}_1$ in Eq. 4 can be stochastically chosen. It is guaranteed that $\underline{C}_1$ will not be not passed over. Only the first ready component in Eq. 4 is a launch component.

Probability current flowing from $C_0$ to $\underline{C}_1$ in Eq. 4 will therefore result in a stochastic hit on $\underline{C}_1$ at some time $t_{sc1}$. When that happens we get the first particle capture together with the next ready component in line.

$$\Psi(t \geq t_{sc1} > t_0) = C_1(t) + \underline{C}_2(t) + \ldots \qquad (5)$$

where $\underline{C}_2(t)$ is zero at $t_{sc1}$. From this point on, second order components such as $\underline{C}_3(t)$ will not be explicitly shown in a q-rule equation because they cannot be launch components. Their presence will be noted by +… following the launch component. These second order components are certainly present in the Schrödinger equation but they cannot be stochastically chosen according to q-rule (2), so they serve no purpose in the equation.

Following Eq. 5 another stochastic hit at time $t_{sc2}$ gives the second particle capture

$$\Psi(t \geq t_{sc2} > t_{sc1} > t_0) = C_2(t) + \underline{C}_3(t) + \ldots \qquad (6)$$

and so fourth. In Eqs. 5 and 6 the correct 'sequential' order of statistical hits is guaranteed by q-rule (2).

It is characteristic of standard quantum theory that there is only one solution to the Schrödinger equation for the given initial conditions, whereas the q-rules provide a separate solution for each quantum jump (Eqs. 5, 6, etc.). The launch component will provide the boundary conditions of the next solution, so $\underline{C}_2(t)$ in Eq. 5 defines the initial boundary of the collapsed solution in Eq. 6.

There is no contradiction between the predictions of the q-rules and standard quantum mechanics. The q-rules are concerned with the probability that a stochastic hit will occur in the next interval $dt$ of time; and standard theory is concerned with the probability distribution of an ensemble of states at some finite time $t$ after the apparatus is turned on. These different rules ask different questions having different answers. However, either one of these protocols can be successfully mapped onto the system, so there can be no observational contradiction.

Equation 4 applies to microscopic states as well, because serial order is just as important in these cases. Atomic states that decay from an initial excited state $a_0$ will go to the next lower energy state $a_1$, and then lower to $a_2$ without skipping a step – unless that possibility is allowed by the Hamiltonian. If it is not allowed, then $a_1$ will not be skipped over any more than $C_1$ in the above macroscopic counter. As in the macroscopic case, q-rule (2) is an essential moderator of any serial sequence at the atomic level. Otherwise, the second order component $a_2$ might be stochastically chosen before $a_1$ and that would be unphysical. As it is, the photon between states $a_0$ and $a_1$ will be released before the photon between state $a_1$ an $a_2$, and there will be no photon between states $a_0$ and $a_2$. Although the q-rules are empirically discovered by investigating macroscopic systems, they can be extended to this microscopic system, thereby supporting the claim that the q-rules apply independent of size.

## Parallel discontinuities

The q-rules also correctly describe macroscopic 'parallel branching' of counters that produce many cascade photons with each capture of a radioactive particle. Imagine two side-by-side counters that are exposed to a radioactive source and are represented by the q-rule equation

$$\Psi(t \geq t_0) = C_0(t) + \underline{C_r}(t) + \underline{C_l}(t) + \ldots \qquad (7)$$

where the initial component $C_0(t)$ means that neither counter has yet captured a particle, $C_r(t)$ means that the counter on the right is the first to make a capture, and $C_l(t)$ means that the counter on the left is the first to make a capture. Let each counter turn off after a capture. Again, we simplify by not including the particle fields.

The ready components $\underline{C}_r(t)$ and $\underline{C}_l(t)$ are initially equal to zero and increase in time. Each one receives probability current from the first component that makes each a direct candidate for a state reducing stochastic hit. Each is a launch component, where $\underline{C}_r(t)$ is the initial boundary conditions for a launch to the right and $\underline{C}_l(t)$ is the initial boundary conditions for a launch to the left.

If the launch component $\underline{C}_r$ in Eq. 7 is stochastically chosen at time $t_{scr}$, the resulting state reduction will be

$$\Psi(t \geq t_{scr} > t_0) = C_r(t) + \underline{C}_f(t)$$

where $\underline{C}_f(t)$ is the launch component to the final state of the system in which both counters have captured a particle. The final ready component $\underline{C}_f(t)$ is not shown in Eq. 7 because it is a second order transition. It cannot be chosen before one of the intermediate components is chosen.

When it is stochastically chosen at time $t_{scf}$ the system will be in its final state

$$\Psi(t \geq t_{scf} > t_{scr} > t_0) = C_f(t).$$

This sequence will follow the left-hand path if the launch component $\underline{C}_l(t)$ in Eq. 7 is stochastically chosen before $\underline{C}_r(t)$ is chosen.

The second q-rule therefore has the effect of forcing these macroscopic counters to go along a right or left path to the final state. Without the second q-rule a second order transition might skip over the intermediate components to score a direct stochastic hit on $C_f$, without one of the intermediate component being definitely involved. This is unphysical. Here again we see the indispensability of q-rule (2) if macroscopic objects are to be quantum mechanically described with this protocol.

The same will be true of *microscopic* parallel systems. Any non-periodic, quantum jump imposes an abrupt and lasting change of a distinctive kind in some part of the universe – even in a microscopic case. Using atomic instead of counter variables, let Eq. 7 represent two alternative routes from a high-energy atomic state $a_0$ to the ground state $a_f$, where the intermediate components do not interact with each other. Each step along the way creates a photon that qualifies it as a quantum jump. The two photons that are released along each path will leave an indelible record that will be different for each path (assuming non-degeneracy); so if the two photons associated with one path are found in the wider universe, then that path must have been stochastically chosen. It is not possible for *all four* photons to be found in a single trial. It will be either the two photons from the left or the two from the right. The released photons are the abrupt and lasting change referred to above, and their distinctive characteristics along each path removes any doubt as to which pathway is finally traversed.

More generally for any microscopic/macroscopic, series/parallel combination of paths, any single path segment that follows or precedes a non-periodic, quantum jump can be correctly described by a q-rule equation.

**Add an observer**

Imagine that an observer witnesses the capture of the particle *p* in Eq. 2.

$$\Psi(t \geq t_0) = pd_0B_0(t) + \underline{d_1}B_1(t) \qquad (8)$$

where $B_0$ is the brain of a conscious observer witnessing the detector $d_0$ in its ground state, and $B_1$ is the brain of the observer witnessing the detector $d_1$ in its capture state. As before, it is not necessary to underline more than one state in the ready component. Because $B_1$ is in the ready component it is not yet empirically realized, so it cannot be a conscious brain. Until there is a stochastic hit on $\underline{d_1}B_1$ the observer is only conscious of the detector in its ground state through the conscious state $B_0$.

Probability current flowing from the first to the second component in Eq. 8 may produce a stochastic hit resulting in

$$\Psi(t \geq t_{sc} > t_0) = d_1 B_1(t) \tag{9}$$

so the brain state $B_1$ becomes part of a realized component at time $t_{sc}$, which means that the observer becomes consciously aware of the capture at that time.

Equation 8 is also correct in standard quantum mechanics. However, in standard theory the second component in Eq. 8 would be said to have the same empirical significance as the first. When applied to an individual trial (i.e., not an ensemble of trials), this produces a paradoxical situation reminiscent of Schrödinger's cat experiment. The brain of the observer would then be seen to be consciously observing the detector in both its $d_0$ and its $d_1$ state at the same time.

This difficulty is related to the fact that standard theory regards Eq. 8 as a complete dynamic process, whereas the q-rules include Eq. 8 and Eq. 9 as separate processes. Typically, standard theory employs only one set of boundary conditions (i.e., the initial conditions), whereas the q-rules employ multiple boundary conditions – two in this case. Every measurement introduces new boundaries, so the initial boundary conditions of Eq. 8 are supplemented by new boundary information to the effect that the particle has been captured – giving the initial conditions of Eq. 9. Standard theory fails to ground the Schrödinger equation in new boundaries when they occur, whereas q-rule theory recognizes new boundary information every time there is a collapse of the wave.

It is possible to refine the account described in Eqs. 8 and 9. To this end, the initial detector $d_0$ is understood to mean the laboratory apparatus *plus* the physiology of the observer up to that part of the brain that records conscious experiences. The detector then includes all the brain parts that are engaged in image processing prior to conscious experience, and the brain state is confined to the part of the cerebral physiology (i.e., presumably the neocortex) that supports conscious experiences.

We then write Eq. 8 in a more limited form

$$\Psi(t \geq t_0) = pd_0B_0(t) + \underline{d}_{w1}B_0(t) \tag{10}$$

where $d_{w1}$ is just the window of the detector that the particle first enters. When the launch component $\underline{d}_{w1}B_0(t)$ is stochastically chosen, the result is followed by a continuous progression of the signal through the detector that now includes the data processing part of the brain, up to and including the conscious part of the brain.

$$\Psi(t \geq t_{sc} > t_0) = d_{w1}B_0(t) \rightarrow d_{i1}B_0(t) \rightarrow d_{f1}B_1(t) \tag{11}$$

where the arrows represent a continuous progression. There are a great may quantum jumps taking place beneath the macroscopic surface of this equation; however, we ignore this detail and focus on the continuous macroscopic appearance of things. The three terms in Eq. 11 then represent a single realized component that evolves continuously in time under the Schrödinger equation. State $d_{i1}$ is the detector when the signal has reached the half-way mark, and $d_{f1}$ is the detector when the signal has finally reached the neocortex, at which time the brain $B_1$ will be conscious of the detector in its capture state.

It may appear that we have revived a cat-like paradox because Eq. 11 contains both the conscious pre-capture brain state $B_0$ and the conscious post-capture state $B_1$. However, the equation does not include these two states at the same time so a paradox is avoided. The q-rules therefore allow a secondary observer to be admitted to the system without a cat-like ambiguity of the kind that concerned Schrödinger. But more than that, the q-rules allow the primary observer to include himself in the system. He has only to imagine that it is his brain that is in contact with the detector, and the Schrödinger dynamics will predict his experience. In this respect, the relationship of the primary observer to the system under the q-rules is similar to that in classical physics.

As before, a more refined description is required because Eqs. 10 and 11 are still not quite right. A fully correct account is possible only when the process given in the Appendix is applied to this case, resulting in many faux detections, or unreal detections, occurring before the real one. The 'observed detector' then follows a pattern similar to that of the 'detector capture' and the 'free neutron decay' that are described the Appendix.

## *Compton scattering*

The Compton scattering of a photon off of an electron center is surely non-periodic. It might appear to represent a quantum discontinuity because the initial momentum is carried into a scattered momentum in a way that involves Planck's constant. However, there are no new particles created or destroyed during the process, so there is no ready component or quantum jump and no collapse of the wave. The q-rule equation is given by $\Psi = \gamma e$ = constant, where $\gamma$ is the incoming photon and $e$ is the electron. Although the two parts of $\Psi$ undergo dramatic continuous and correlated change when the space and time variables are included, they have no effect on the magnitude of the "trans-representational" q-rule component $\gamma e$. The q-rule equation is therefore constant. The same may be said of Bragg scattering.

## **Atomic absorption and emission**

Applying this scheme to the case of atomic absorption and emission, the atom $a$ in its ground state interacts with a laser field $\gamma_N$ containing $N$ photons. These photons have a frequency 0-1, where 0 refers to the ground state $a_0$ and 1 refers to the excited state $a_1$. In addition, a photon is spontaneously released from the excited state atom giving the q-rule equation.

$$\Psi(t \geq t_0) = \gamma_N a_0 \Leftrightarrow \gamma_{N-1} a_1 + \gamma_{N-1} \underline{a}_0 \otimes \gamma \tag{12}$$

where only the first component is non-zero at time $t_0$. Each component in this equation is a function of time but that is not specifically shown in order to simplify the notation. The double arrow ($\Leftrightarrow$) represents a stimulated oscillation that begins at $t_0$. The atom oscillates between its ground state and the exited state. In its ground state it absorbs a photon from the laser beam, and in its capture state it is stimulated to emit a photon to the laser beam. And finally, when the atom is in the capture state a spontaneous emission to ground becomes a possibility. That emission is non-periodic and the resulting component $\gamma_{N-1}\underline{a_0}\otimes\gamma$ is a quantum jump that creates the photon $\gamma$.

If the atom begins in the excited state and is exposed to a laser beam, we get

$$\Psi(t \geq t_0) = \gamma_N\, a_1 \Leftrightarrow \gamma_{N+1} a_0 + \gamma_N\, \underline{a_0}\otimes\gamma \qquad (13)$$

where again, only the first component is non-zero at $t_0$. It does not matter if the oscillation begins in $a_0$ or in $a_1$.

## A laser

Given a four-level atom with a ground state $a_0$ and three excited states $a_1, a_2, a_3$ of increasing energy. It is immersed in a laser field of $N$ photons with energy connecting levels $a_1$ and $a_2$. The atom is initially pumped into the short-lived state $a_3$ and quickly drops to $a_2$ with an energy loss that may involve some dissipative process $e_x$ like molecular collisions.

$$\Psi(t \geq t_0) = \gamma_N a_3 + \gamma_N a_2 \otimes e_x \qquad (14)$$

where the second (metastable) component is zero at $t_0$ and increases in time. This is written as a two component q-rule equation although there is no quantum jump involved. Again, the explicit time dependence of the components is not shown. A subsequent stimulated oscillation is followed by another loss of energy given by $e_{xx}$.

$$\begin{array}{ccc} \text{metastable} & & \text{short-lived} \\ \Psi(t \geq t_0) = & \gamma_N a_2 \otimes e_x \quad \Leftrightarrow \quad \gamma_{N+1} a_1 \otimes e_x + \gamma_{N+1} a_0 \otimes e_x \otimes e_{xx} \end{array} \quad (15)$$

The double arrow again represents the stimulated oscillation that is due to the presence of a laser beam connecting these levels. The last component $\gamma_{N+1}a_0 \otimes e_x \otimes e_{xx}$ in this equation is short lived, but we are going to assume that the energy loss $e_{xx}$ is not due to radiation so we will not treat the component as being ready. There are as yet no quantum jumps following the initial pumping action. However, as soon as the state $a_0$ is produced it is exposed to a pumping action photon $\gamma'$ giving

$$\Psi(t \geq t_0) = \gamma' \gamma_{N+1} a_0 \otimes e_x \otimes e_{xx} + \gamma_{N+1} \underline{a}_3 \otimes e_x \otimes e_{xx}$$

A stochastic hit at time $t_{sc}$ will collapse the first component in this equation as well as all the left-over components from Eqs. 14 and 15, leaving just

$$\Psi(t \geq t_{sc} > t_0) = \gamma_{N+1} a_3 \otimes e_x \otimes e_{xx}$$

This is the same as the original state $\gamma_N a_3$ in Eq. 14 except that it includes an additional photon in the laser beam and some dissipative terms that carry over from the previous cycle. Comparing the original state $\gamma_N a_3$ with the final state $\gamma_{N+1} a_3 \otimes e_x \otimes e_{xx}$, it is clear that the energy absorbed from the pumping source is equal to the energy of the new photon in the laser beam plus the two dissipative processes $e_x$ and $e_{xx}$. This cycle is repeated many times resulting in pumping many new photons into the laser beam. Evidently each photon pumped into the beam involves just one stochastic hit.

## Other applications

In another paper the q-rules are applied to the problem of how a widespread quantum mechanical wave function can be localized [1]. Optical shelving is an unusual phenomenon that is not easy to understand without a q-rule analysis that is given separately [2]. And finally, we propose an experimental test of the q-rules that distinguishes them from the standard understanding and from the GRW/CSL theory of Ghirardi and Pearle [3].

## Appendix – faux processes

Let a non-periodic interaction give rise to a ready component in the equation

$$\Psi(t \geq t_0) = \psi_0(t) + \int_0^{a(t-t_0)} \underline{\psi}_1(t,\tau)d\tau \tag{16}$$

where the space coordinates have been integrated out. It is assumed that the ready component $\underline{\psi}_1$ is a quantum jump away from $\psi_0$ in that it contains new particles or annihilates old ones. The constant *a* has units of inverse time. Some of the variables in this integral evolve in time *t* independent of a stochastic choice, and others vary with the unitless parameter $\tau$ that is a function of stochastic choice. At every moment of time the Schrödinger equation initiates a 'possible' evolution that the second component in Eq. 16 carries to completion as though it had actually occurred. The integral is therefore made up of ready components representing many *faux processes* that are not empirically real. It is a sum of all possible evolutions coming off of $\psi_0$, where $\tau_0$ is set equal to zero at the onset of each one.

Starting at time $t_0$ and skipping to finitely separated times $t_1$, $t_2$, $t_3$, etc., the resulting integrand as a function of *t* and $\tau$ is equal to

*$t_0$* : **$\underline{\psi}_1(t_0, \tau_0)d\tau$**

*$t_1$* : $\underline{\psi}_1(t_1, \tau_0)d\tau$ + **$\underline{\psi}_1(t_1, \tau_1)d\tau$**

*$t_2$* : $\underline{\psi}_1(t_2, \tau_0)d\tau$ + $\underline{\psi}_1(t_2, \tau_1)d\tau$ + **$\underline{\psi}_1(t_2, \tau_2)d\tau$**  (17)

*$t_3$* : $\underline{\psi}_1(t_3, \tau_0)d\tau$ + $\underline{\psi}_1(t_3, \tau_1)d\tau$ + $\underline{\psi}_1(t_3, \tau_2)d\tau$ + **$\underline{\psi}_1(t_3, \tau_3)d\tau$**

*$t_4$* : $\underline{\psi}_1(t_4, \tau_0)d\tau$ + $\underline{\psi}_1(t_4, \tau_1)d\tau$ + $\underline{\psi}_1(t_4, \tau_2)d\tau$ + $\underline{\psi}_1(t_4, \tau_3)d\tau$ + **$\underline{\psi}_1(t_4, \tau_4)d\tau$**

*$t_5$* : *etc.*

where the intervals between $t_0$, $t_1$, $t_2$, etc. are really a continuum of infinitesimal intervals in the variable $t$.

The ready component $\underline{\psi}_1(t_0, \tau_0)d\tau$ at time $t_0$ is advanced by the Schrödinger equation along the diagonal of bold face components in the array. This diagonal shows the evolution of the process that is begun at $t_0$ and proceeds independent of other processes that are initiated at other times. A new process begins at each moment of time because the first component $\psi_0$ in Eq. 16 continuously feeds current into the second component. Continuity of the Schrödinger equation requires that the state $\underline{\psi}_1$ at the head of each diagonal is functionally identical with $\psi_0$ and becomes the new function $\underline{\psi}_1$ only after moving down the diagonal for a time $\Delta T$. This is difficult to display in Eq. 16 or in the array of Eq. 17, so we just state it as something to remember when considering this equation and this array. When writing these equations and other q-rule equations we generally ignore the transition time $\Delta T$ and the associated uncertainty in energy.

It is the interaction that produces the array in Eq. 17 from the start. The interaction Hamiltonian initiates the infinitesimal faux process that goes down the diagonal of the array from time $t_0$, and so on down the line. At every moment along the way it generates an infinitesimal process of this kind that is unreal and unrealized, and will continue to do so until the initial component $\psi_0$ has diminished to zero or has otherwise been disengaged. But while the interaction Hamiltonian initiates these processes it cannot itself bring the array to a stochastic conclusion. That is why we need the q-rules – to decide when to collapse all those faux processes and make one of them an empirical reality.

The standard Copenhagen interpretation of quantum mechanics never gets beyond Eq. 17. All the components in the array at time $t_4$ represent possible states of the system at that time, and their total square modulus is the probability that there has been a stochastic hit by that time. Objective reality is not the issue from a Copenhagen point of view. The meaning of these component magnitudes has only to do with the probability of their being observed. But from a q-rule point of view observation has nothing to do with a collapse because a stochastic choice is an *objective* choice. Collapse does not depend on the presence or absence of an observer. The empirically unreal process in Eq. 17 is brought to a realistic conclusion by the objectively determined q-rules – not by subjectively assembled ensembles.

Only the first component along any horizontal line of the array is a launch component for only it receives probability current directly from the realized component $\psi_0$. Therefore, only the first component in the array can be stochastically chosen. If that happens at time $t_{sc} = t_5$, then following $t_3$ the array in Eq. 17 will become.

$$t_4: \psi_1(t_4, \tau_0)d\tau + \psi_1(t_4, \tau_1)d\tau + \psi_1(t_4, \tau_2)d\tau + \psi_1(t_4, \tau_3)d\tau + \psi_1(t_4, \tau_4)d\tau$$

$$t_5: \psi_1(t_5, \tau_0)$$

$$t_6: \qquad \psi_1(t_6, \tau_1) \tag{18}$$

$$t_7: \qquad\qquad \psi_1(t_7, \tau_2)$$

$$t_8: \qquad\qquad\qquad \text{etc.}$$

The third q-rule requires that all of the off-diagonal components in Eq. 18 are equal to zero after $t_{sc}$, and that the diagonal component is realized and normalized. The latter is accomplished by removing $d\tau$ from each of the diagonal terms. The first few realized diagonals in this equation, beginning with $t_5$, are subject to the transient uncertainty $\Delta E$ as $\psi_0$ changes *continuously* into $\psi_1$ in time $\Delta T$.

Equation 16 can be written in form in which only launch components are included

$$\Psi(t \geq t_0) = \psi_0(t) + \psi_1(t, \tau_0)d\tau + \ldots$$

and this results in a *realize* component at the time $t_{sc}$ of stochastic choice that equals

$$\Psi(t_{sc} > t_0) = \psi_1(t_{sc}, \tau_0) \tag{19}$$

followed by $\Psi(t \geq t_{sc} > t_0) = \psi_1(t, \tau)$

It must be remembered that the function $\psi_1(t_{sc}, \tau_0)$ in this equation is really $\psi_0(t_{sc}, \tau_0)$ and only become $\psi_1(t, \tau)$ when transients die out in time $\Delta T$.

The consequences of Eq. 19 when dealing with a detector capture and with a neutron decay are illustrated below.

**A particle capture**

The description of a detector that captures a particle develops in two different ways: One is dependent on time $t$ and the other is dependent on the unitless parameter $\tau$. Imagine that the detector contains a clock that is set to read $t_0$ at the beginning of the experiment and ticks continuously thereafter. Its behavior throughout the experiment will proceed without regard to the possibility of capture, so its variables will depend on the time $t$. But the ionic cascade that is initiated when the particle enters the detector's window is different. The time of that event is uncertain before there has been a stochastic hit, so the wave equation will include the "possibility" of a cascade beginning at each moment of time after the interaction begins. These are *faux cascades* because they are not empirically real. They exist only in the ready components of the integral, where each is initiated with the setting $\tau_0 = 0$. For the case of the detector $d_1$, the term $\psi_1(t_k, \tau_0)$ at time $t_k$ in array of Eq. 17 is equal to $\underline{d}_1(t_k, \tau_0)$ and represents the faux capture beginning at that time. A *realized* capture (or cascade) beginning at the time $t_{sc}$ of a stochastic hit is therefore given by the component

$$\Psi(t_{sc} \geq t_0) = d_1(t_{sc}, \tau_0)$$

followed by $\Psi(t \geq t_{sc} > t_0) = d_1(t, \tau)$

**A free neutron decay**

Like a detector capture, a neutron decay is characterized by two times. First there is the metric background time $t$ that describes the progress of the neutron across the laboratory before decay, and continues to describe the progress of the decay products after decay. And second, there is the time $\tau$ that resets to zero at each moment of time.

Consequently, the neutron will spew out faux decay particles in all directions as it moves across the laboratory, where each of these is keyed to the temporal parameter $\tau_0 = 0$ in Eqs. 16 and 17. The term $\underline{\psi}_1(t_k, \tau_0)$ at time $t_k$ in the array of Eq. 17 is then equal to $\underline{ep\bar{v}}(t_k, \tau_0)$ and represents a *faux decay* beginning at that time. Decay particles will not become empirically real until a time $t_{sc}$ when a stochastic hit collapses the integral, thereby *realizing* a decay with the initial conditions given by

$$\Psi(t_{sc} \geq t_0) = ep\bar{v}(t_{sc}, \tau_0)$$

followed by $\Psi(t \geq t_{sc} > t_0) = ep\bar{v}(t, \tau)$

In the q-rule equations appearing in other papers the transient time $\Delta T$ and the associated energy uncertainty $\Delta E$ are routinely ignored.